\documentclass[aps,pra,10pt,twocolumn,superscriptaddress]{revtex4-1}

\usepackage{amsfonts}	% need for mathbb
\usepackage{amsmath}    % need for subequations
\usepackage{amssymb}
\usepackage{bm}
\usepackage{bbold}
\usepackage{graphicx}   % need for figures
\usepackage{verbatim}   % useful for program listings
\usepackage{color}      % use if color is used in text
\usepackage[subrefformat=parens,labelformat=parens,caption=false]{subfig}	% use for side-by-side figures
\usepackage{hyperref}   % use for hypertext links, including those to external documents and URLs
\usepackage{mathtools}

\usepackage{todonotes}

\newcommand{\lef}{\left(}
\newcommand{\rig}{\right)}

\newcommand{\imu}{i}
\newcommand{\re}{\mathrm{Re}}
\newcommand{\im}{\mathrm{Im}}

\newcommand{\deriv}{\mathrm{d}}
\newcommand{\sign}{\mathrm{sgn}}

\newcommand{\vecr}{\mathbf r}

\newcommand{\veck}{\mathbf k}

\newcommand{\vecv}{\mathbf v}
\newcommand{\vecw}{\mathbf w}

\newcommand{\vece}{\mathbf e}

\newcommand{\Ham}{H}
\newcommand{\Green}{\mathcal{G}}
\newcommand{\barGreen}{\bar{\mathcal{G}}}

\newcommand{\Vol}{\mathcal{V}}
\newcommand{\cutoff}{\mathrm{C}}

\newcommand{\LOA}{{\mathrm{(LO)}}}
\newcommand{\FOB}{{\mathrm{(FB)}}}
\newcommand{\SCB}{{\mathrm{(SB)}}}

\begin{document}
		
		\title{Disorder-driven exceptional lines and Fermi ribbons in tilted nodal-line semimetals}
		
		\author{Kristof Moors}
		\email{kristof.moors@uni.lu}
		\affiliation{University of Luxembourg, Physics and Materials Science Research Unit, Avenue de la Fa\"iencerie 162a, L-1511 Luxembourg, Luxembourg}
		\author{Alexander A. Zyuzin}
		\affiliation{Department of Applied Physics, Aalto University, P.\ O.\ Box 15100, FI-00076 AALTO, Finland}
		\affiliation{Ioffe Physical-Technical Institute, 194021 St.\ Petersburg, Russia}
		\author{Alexander Yu. Zyuzin}
		\affiliation{Ioffe Physical-Technical Institute, 194021 St.\ Petersburg, Russia}
		\author{Rakesh P. Tiwari}
		\affiliation{Department of Physics, McGill University, Montr\'eal, Qu\'ebec, Canada H3A 2T8}
		\author{Thomas L. Schmidt}
		\affiliation{University of Luxembourg, Physics and Materials Science Research Unit, Avenue de la Fa\"iencerie 162a, L-1511 Luxembourg, Luxembourg}
		
		\date{\today}
		
		\begin{abstract}
			We consider the impact of disorder on the spectrum of three-dimensional nodal-line semimetals.
			We show that the combination of disorder and a tilted spectrum naturally leads to a non-Hermitian self-energy contribution that can split a nodal line into a pair of exceptional lines.
			These exceptional lines form the boundary of an open and orientable bulk Fermi ribbon in reciprocal space on which the energy gap vanishes.
			We find that the orientation and shape of such a disorder-induced bulk Fermi ribbon is controlled by the tilt direction and the disorder properties, which can also be exploited to realize a twisted bulk Fermi ribbon with nontrivial winding number.
			Our results put forward a paradigm for the exploration of non-Hermitian topological phases of matter.
		\end{abstract}
		
		\maketitle
		
		\textit{Introduction}.---Recently, there has been a growing interest in non-Hermitian topological phases of matter (see Ref.~\cite{MartinezAlvarez2018} for a recent overview).
		A Hamiltonian with non-Hermitian terms can be considered to represent, e.g.,
		contact with the environment (open quantum systems \cite{Carmichael1993, Rotter2011}),
		dissipation and driving (e.g., resonator circuits and photonics or atomic gas systems with gain and loss \cite{Stehmann2004, Longhi2010, Regensburger2012, Malzard2015, Xu2017, Zhou2018, Cerjan2018A}).
		Another scenario to realize non-Hermitian systems is to reinterpret the complex self-energy of quasiparticles with a finite lifetime due to certain interactions or disorder on a microscopic level (e.g., electrons subject to electron-phonon and electron-electron interactions or impurities \cite{Kozii2017, Zyuzin2018, Papaj2018, Yoshida2018}) as the non-Hermitian term of an effective Hamiltonian.
		Given their profound implications and vast applicability, the question naturally arises whether the concepts that underlie the classification of topological phases of matter (e.g., topological invariants, the bulk-boundary correspondence, topology- and symmetry-protected bulk or surface states) can be extended to non-Hermitian systems. How and to what extent this can be done is a topic of active research \cite{Hu2011, Esaki2011, Liang2013, Yuce2015, Rudner2016, Lee2016, Leykam2017, Lieu2018, Gong2018, Xiong2018, Shen2018, Yao2018A, Kawabata2018, Yin2018, Kunst2018, Dangel2018, Zhang2018A, Zhang2018B, Yao2018B, Jin2018}.
		
		In Hermitian systems, the notions of a gapped phase and band touchings are crucial for the bulk-boundary correspondence and topologically protected states.
		These notions do not carry over to the non-Hermitian case in a straightforward manner, complicating the search for well-behaved topological invariants that could take over the role of invariants in Hermitian systems, for example the Chern number.
		There is no unique way to extend the notion of a gapped phase for a complex quasiparticle spectrum \cite{Gong2018, Shen2018}, and band touchings can turn into exceptional points where the Hamiltonian is defective \cite{Berry2004, Heiss2012, MartinezAlvarez2018}. The latter implies that a complete set of eigenvectors cannot be retrieved at these points.
		
		For non-Hermitian systems with two bands in two (2D) and three (3D) spatial dimensions, generic band touchings typically lead to exceptional points and lines \cite{Berry2004}, respectively.
		It was shown that an exceptional loop can be obtained from a Weyl node by adding non-Hermitian terms to the Hamiltonian \cite{Xu2017, Cerjan2018A, Zyuzin2018}. This was recently identified in an optical waveguide \cite{Cerjan2018B}.
		Exceptional lines also form the termination lines of zero-energy surfaces \cite{Zyuzin2018, Carlstrom2018A}.
		Such zero-energy surfaces form the natural generalization of bulk Fermi arcs in 2D non-Hermitian systems \cite{Kozii2017} (confirmed experimentally in photonics crystal slabs \cite{Zhou2018}) and have been dubbed \textit{bulk Fermi ribbons} \cite{Carlstrom2018A}.
		Furthermore, exceptional lines can form knots and links (forming so called \textit{knotted non-Hermitian metals} as introduced in Ref.~\cite{Carlstrom2018B}) such that the associated Fermi ribbons generally form closed and orientable surfaces with a possibly nontrivial topology, which can generally be classified as Seifert surfaces \cite{Seifert1935, Carlstrom2018A}.
		
		Topological nodal-line semimetals (NLSMs) are systems that form robust line-like band touchings in the Hermitian regime \cite{Burkov2011, Fang2015, Bian2016, Hu2016, Fang2016}. Recently, it was shown that these nodal lines can be split into two exceptional lines, connected by a Fermi ribbon, by adding a non-Hermitian term to the Hamiltonian, representing some form of particle gain and loss in the system \cite{Yang2018, WangH2018}.
		
		In this Rapid Communication, we show that the formation of a pair of exceptional lines and associated bulk Fermi ribbon can occur naturally in 3D NLSMs due to the presence of disorder (e.g., impurities).
		It is the complex self-energy correction to the Green function due to disorder that renders the matrix structure of the quasiparticle's Green function non-Hermitian and, under certain conditions, induces this formation.
		We show that, for a nodal ring with tilted spectrum, disorder leads to the separation of the nodal ring into a pair of exceptional rings, along with the formation of a circular bulk Fermi ribbon.
		Similar to disordered Weyl semimetals \cite{Zyuzin2018}, this phenomenology is most pronounced in case of strong tilt that induces a type-II NLSM.
		Several compounds, such as K${}_4$P${}_3$ and Mg${}_3$Bi${}_2$ \cite{Li2017, Chang2017, WangB2018}, have already been identified as 2D and 3D type-II NLSMs, and they can also be engineered with optical lattices \cite{He2018}.
		We further show the impact of the tilt and disorder properties on the configuration and shape of the exceptional lines and bulk Fermi ribbon and also discuss the possibility of obtaining twisted ribbons with nontrivial winding number.
		
\begin{figure*}[htb]
			\centering
			\subfloat[\ \label{fig:band_structure_a}]{\includegraphics[height=0.18\linewidth]{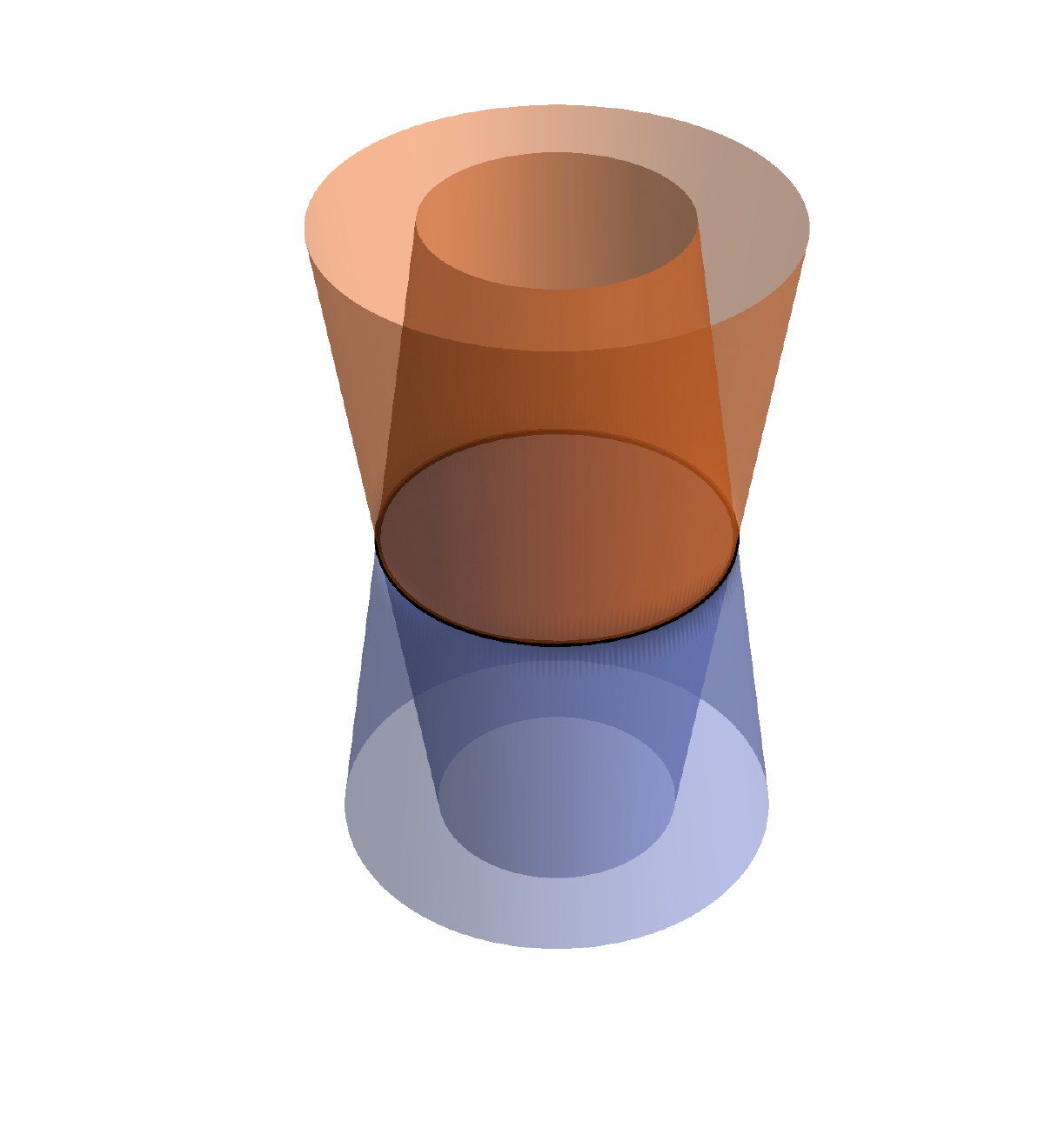}}
			\hspace{0.0\linewidth}
			\subfloat[\ \label{fig:band_structure_b}]{\includegraphics[height=0.18\linewidth]{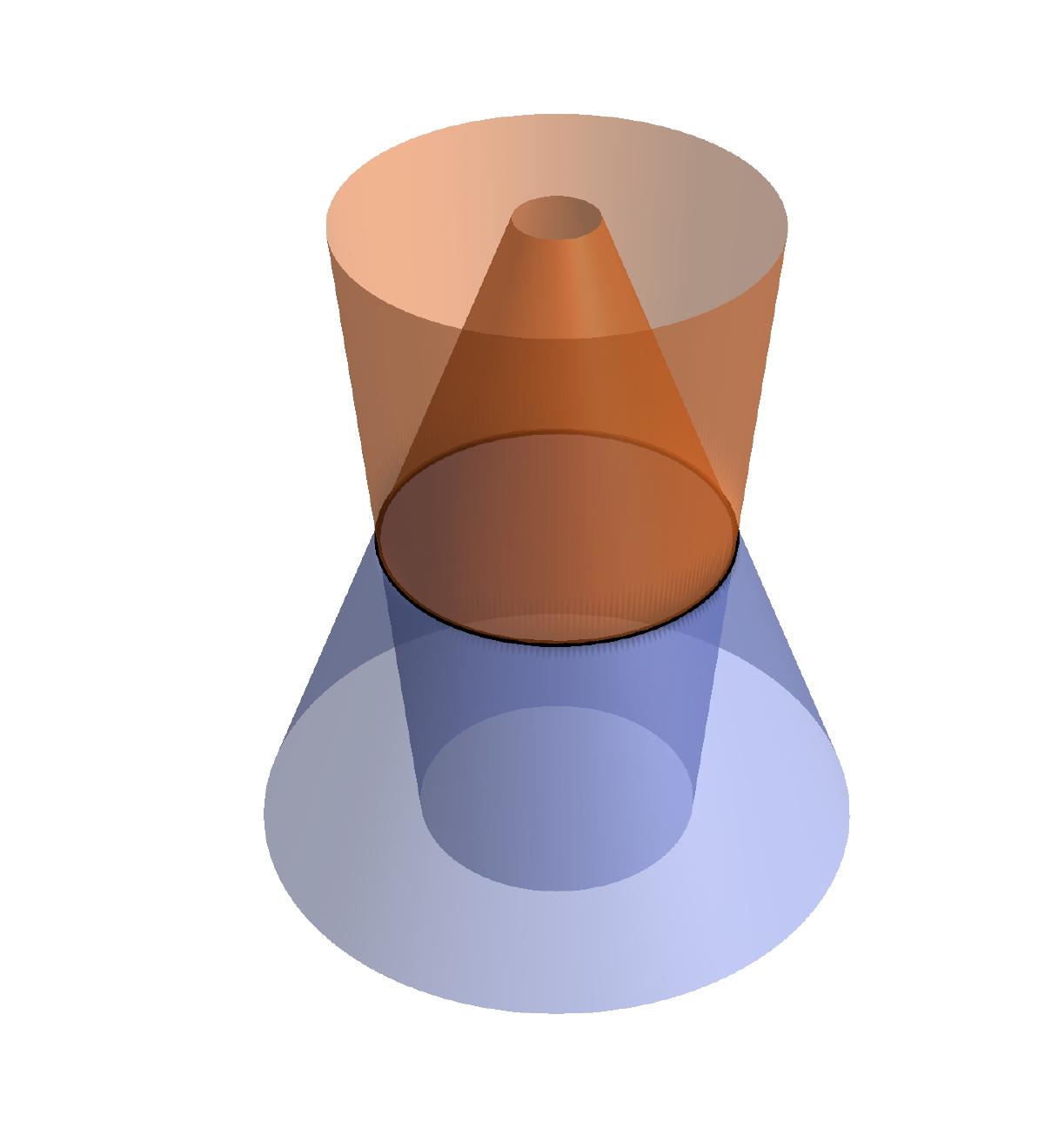}}
			\hspace{0.0\linewidth}
			\subfloat[\ \label{fig:band_structure_c}]{\includegraphics[height=0.18\linewidth]{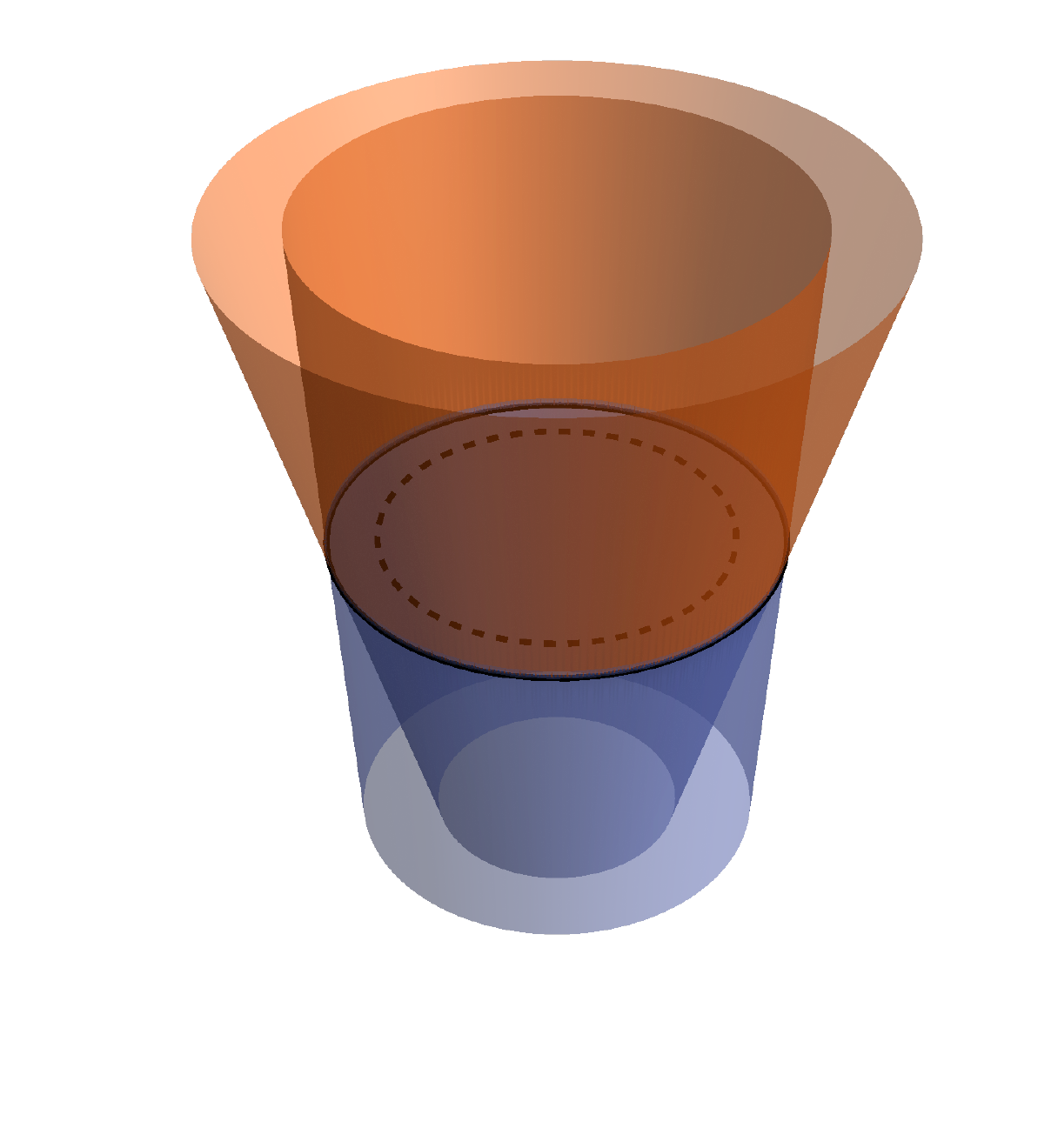}}
			\hspace{0.0\linewidth}
			\subfloat[\ \label{fig:band_structure_d}]{\includegraphics[height=0.18\linewidth]{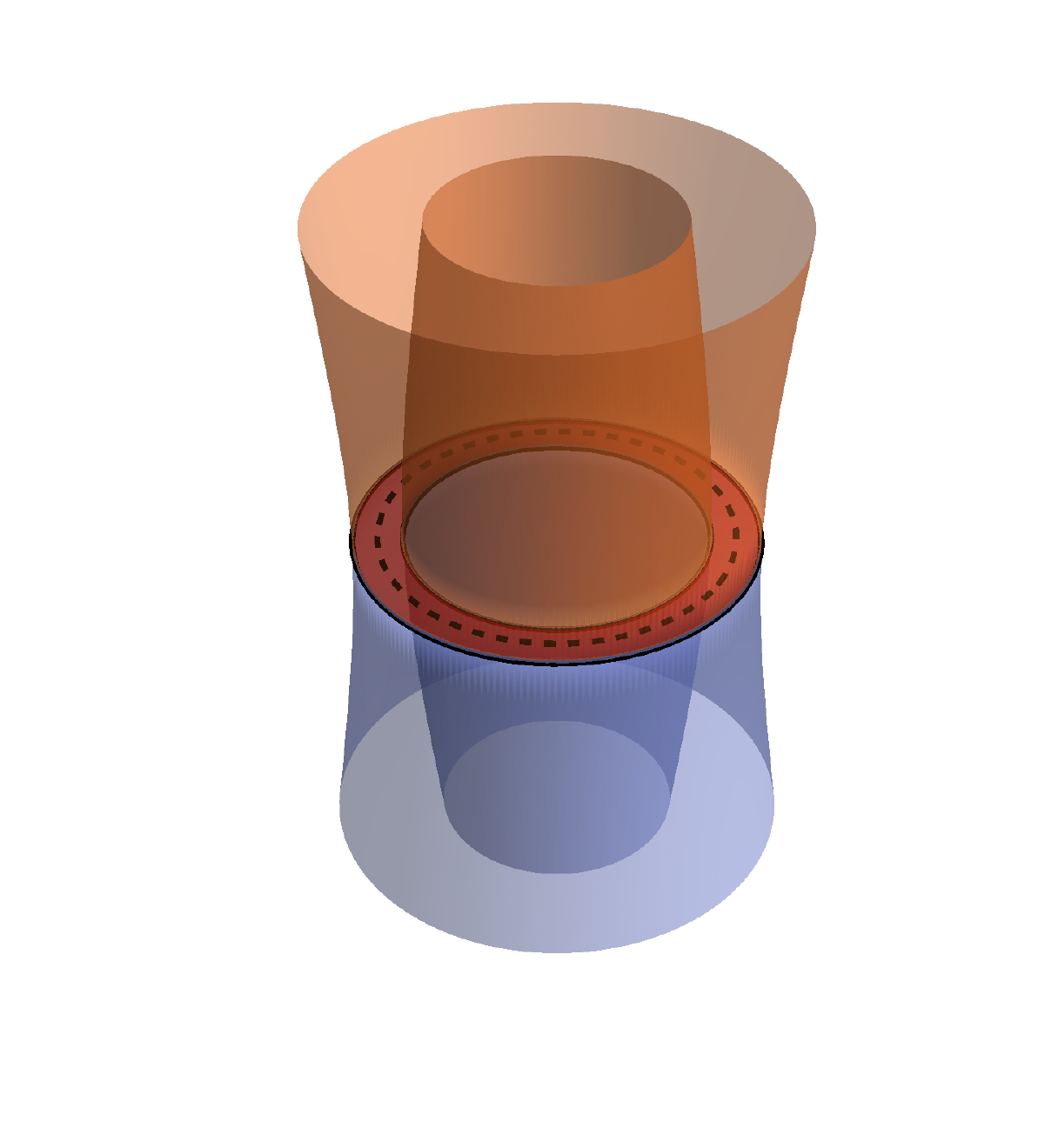}}
			\hspace{0.0\linewidth}
			\subfloat[\ \label{fig:Fermi_ribbons}]{\includegraphics[height=0.18\linewidth]{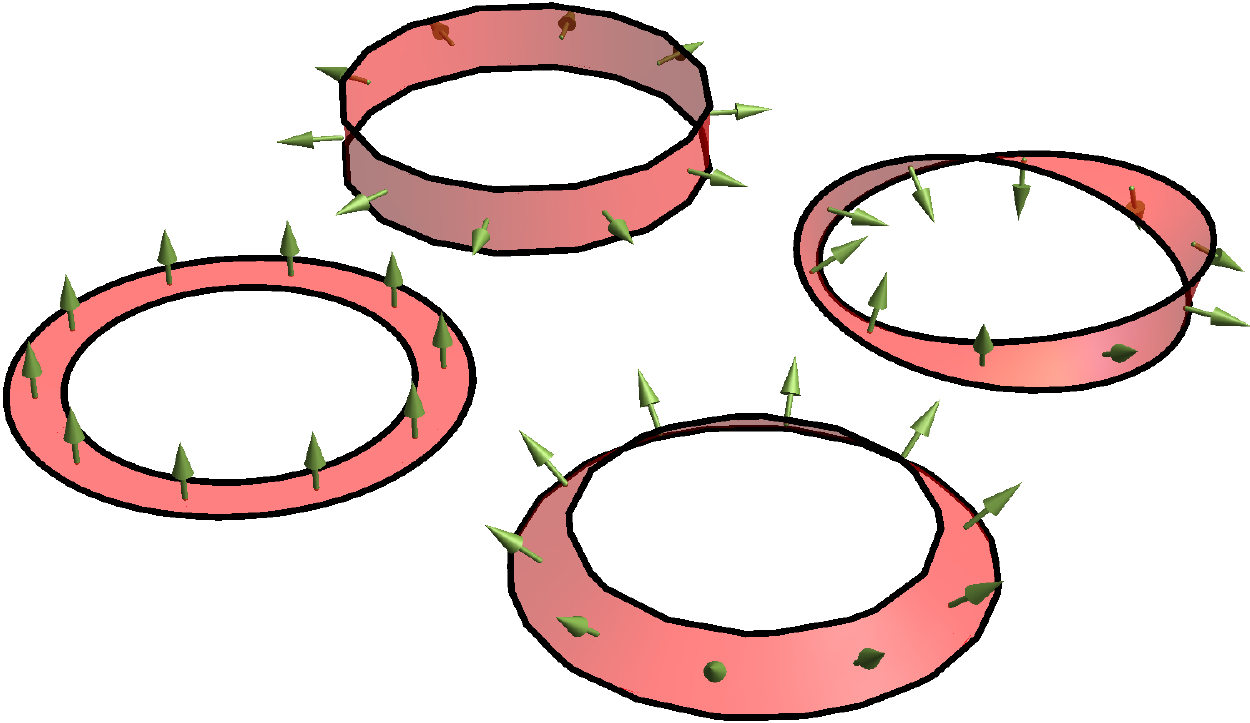}}
			\caption{
				(a)-(d) The real part of the energy spectrum above (orange) and below (blue) the nodal-ring energy is presented as a function of $k_x$ and $k_y$ (with $k_z = 0$), according to Eqs.~\eqref{eq:nodal_loop_spectrum}, \eqref{eq:SE_FOB}, \eqref{eq:SE_FOB_weak_tilt}, and \eqref{eq:SE_FOB_strong_tilt}, for
				a type-I nodal-line semimetal (NLSM) (a) without tilt and (b) with radial tilt,
				a type-II NLSM with (c) radial tilt and (d) axial tilt with first-order Born self-energy correction due to disorder.
				(e) The bulk Fermi ribbons of a disordered type-II NLSM with (left) axial tilt, (top) radial tilt, (bottom) axial and radial tilt, and (right) a twisting tilt vector with winding number equal to one.
				The nodal ring and exceptional lines are indicated in black, the bulk Fermi ribbon in red, and the tilt direction by the green arrows.
			}
			\label{fig:band_structure}
		\end{figure*}

		\textit{Tilted nodal-line semimetals}.---We proceed with an analytically tractable NLSM, considering a nodal ring with radius $Q$ in the $k_z=0$ plane, centered around the origin, with the following two-band Hamiltonian:
		\begin{align} \label{eq:nodal_loop_Ham}
		\begin{split}
		\Ham(\veck) &= [\hbar w_R (k_R - Q) + \hbar w_z k_z] \, \sigma_0 \\
		&\quad + \hbar v_R (k_R - Q) \, \sigma_1 + \hbar v_z k_z \, \sigma_3,
		\end{split}
		\end{align}
		with $\veck \equiv (k_x, k_y, k_z)$, $k_R \equiv \sqrt{k_x^2 + k_y^2}$, $\sigma_0 \equiv \mathbb{1}_{2 \times 2}$, and $\sigma_{1,2,3}$ the three Pauli matrices. The parameters $v_R$, $v_z$ determine the slope of the linearized spectrum at the nodal ring in the radial and axial direction, respectively, while $w_R$ and $w_z$ represent a tilt of the spectrum along the respective directions. This leads to the following spectrum [see Fig.~\subref*{fig:band_structure_a}]:
		\begin{align} \label{eq:nodal_loop_spectrum}
		\begin{split}
		E_s(\veck) &= \hbar \vecw \cdot (\veck - Q \, \vece_{k_R}) \\
		&\; \; \; + s \hbar \sqrt{\sum\nolimits_\alpha [\vecv_\alpha \cdot (\veck - Q \, \vece_{k_R}) ]^2} \, ,
		\end{split}
		\end{align}
		with $s = \pm$, $\vecv_1 = (v_R \cos k_\phi, v_R \sin k_\phi, 0)$, $\vecv_2 = (0,0,0)$, $\vecv_3 = (0,0,v_z)$, $\vecw = (w_R \cos k_\phi, w_R \sin k_\phi, w_z)$, $\vece_{k_R}$ the unit vector along the radial direction with respect to the nodal ring and summation over $\alpha = 1,2,3$.
		The nodal ring can be protected by one or several symmetries \cite{Fang2016}.
		
		Similar to the case of a pointlike Weyl node \cite{Soluyanov2015, Deng2016}, the linear spectrum around a nodal line can be strongly tilted such that the slopes of both bands have equal signs, giving rise to a type-II NLSM with electron and hole pockets at the nodal-line energy \cite{Burkov2011, Chang2017, WangB2018, He2018}. From Eq.~\eqref{eq:nodal_loop_spectrum}, we see that the Fermi surface wave vectors for band $s$ at the nodal-ring energy satisfy the following constraints:
		\begin{equation}
		\begin{split}
		&s [w_R (k_R - Q) + w_z k_z] \leq 0, \\
		&k_R = Q + w_R w_z k_z / (v_R^2 - w_R^2) \\
		&\qquad \; \pm \sqrt{k_z^2 (v_R^2 w_z^2 + w_R^2 v_z^2 - v_R^2 v_z^2 )} / (v_R^2 - w_R^2).
		\end{split}
		\end{equation}
		An overtilted spectrum with solutions for $k_z \neq 0$ is realized when $w_R^2/v_R^2 + w_z^2 / v_z^2 > 1$. In this case, the density of states (DoS) close to the nodal ring is equal to $Q k_\cutoff / |\vecw|$ up to a constant prefactor, where $k_\cutoff > 0$ is a UV cutoff wave vector, denoting the maximal distance from the nodal ring along the radial and axial directions (assuming $k_\cutoff < Q$) and required to regularize a fully linearized spectrum (see Sec.~\ref{sec:FBA} in Supplemental Material for details and discussion on the cutoff dependence). Hence, the DoS is finite at the nodal-ring energy.
		An example of a NLSM spectrum with radial tilt is shown in Fig.~\subref*{fig:band_structure_c}.
		
		\textit{Self-energy due to disorder}.---We consider a disorder potential $V(\vecr)$, assuming the potential to originate from $N$ identical localized impurity potentials with Dirac-delta profile at different positions $\vecr_i$ and disorder strength $S_0$:
			\begin{equation} \label{eq:disorder}
			V(\vecr) \equiv \sum_{i=1}^N S_0 \, \delta (\vecr - \vecr_i).
			\end{equation}

		The disorder-averaged retarded Green function $\barGreen(\veck, \omega)$ (conserving wave vector $\veck$) can be obtained by averaging over all the impurity positions with a uniform distribution, leading to
		\begin{equation}
		\langle \Green(\veck, \veck'; \omega) \rangle_{\vecr_i} = \delta_{\veck, \veck'} \, \barGreen(\veck, \omega),
		\end{equation}
		and the following Dyson equation:
		\begin{equation}
		\barGreen(\veck, \omega) = \Green^{(0)}(\veck, \omega) + \Green^{(0)}(\veck, \omega) \Sigma(\veck, \omega) \barGreen(\veck, \omega),
		\end{equation}
		with retarded self-energy $\Sigma(\veck, \omega)$ (in units of frequency).
		
		Within the lowest-order (LO) approximation \cite{Bruus2004}, the spectrum merely shifts in energy by $n S_0 / \hbar$, with $n \equiv N/\Vol$, and the effective Hamiltonian that can be associated to the Green function's quasiparticle spectrum, $\Ham(\veck) + \hbar \Sigma^\LOA$, remains Hermitian. Turning to the first-order Born (FB) approximation of the self-energy, we obtain the following self-energy:
		\begin{equation} \label{eq:SE_FOB}
		\Sigma^\FOB(\omega) = \frac{n S_0^2}{\Vol \hbar^2} \sum_\veck \Green^{(0)}(\veck, \omega).
		\end{equation}
		The resulting complex quasiparticle spectrum $E_s^\FOB(\veck)$ can be obtained from the following equation:
		\begin{align} \label{eq:SE_FOB_eq}
		\det\{ E_s^\FOB - [ H(\veck) + \hbar \Sigma^\FOB(E_s^\FOB/\hbar) ] \} = 0.
		\end{align}
		The consideration of fully localized potentials simplifies the summation over $\veck$, but it does not change the matrix structure and $\omega$ dependence of the self-energy qualitatively as compared to extended potential profiles. Hence, the results and phenomenology below are also relevant for more generic disorder.
		
		First, we consider a type-I NLSM with nodal ring (in the $k_z = 0$ plane with radius $Q$, centered around the origin), isotropic linear dispersion relation ($v_R = v_z \equiv v$) and tilt along $k_z$ determined by $w_z$.
		In the limit of weak axial tilt ($|w_z| \ll |v|$) and close to the nodal-ring energy ($|\omega/v| \ll Q$), we can proceed analytically and the self-energy is given by:
		\begin{align} \label{eq:SE_FOB_weak_tilt}
		\begin{split}
		&\Sigma^\FOB(\omega) \approx
		- \frac{\gamma k_\cutoff^2}{8 \pi v} \, \sigma_1 \\
		&\; \quad
		- \frac{\gamma Q}{4 \pi v^2} \lef 2 \omega \ln\left| \frac{v k_\cutoff}{\omega} \right| + \imu \, \pi |\omega| \rig \lef \sigma_0 - \frac{w_z}{v} \, \sigma_3 \rig,
		\end{split}
		\end{align}
		with $\gamma \equiv n S_0^2/\hbar^2$ and only keeping the leading-order contributions as a function of $\omega$, $k_\cutoff$, and $w_z/v$. This result is valid with chemical potential equal to zero and is easily generalized by the following substitution on the right-hand side: $\omega \rightarrow \Omega \equiv \omega + \mu/\hbar$, with chemical potential $\mu$. Note that this self-energy correction renormalizes the radius of the nodal ring as $Q \rightarrow Q + \gamma k_\cutoff^2/(8 \pi v^2)$, irrespective of the tilt.
		
		The imaginary part of the self-energy leads to a non-Hermitian contribution in the effective Hamiltonian $\Ham(\veck) + \hbar \Sigma^\FOB[E_s^\FOB(\veck)/\hbar]$. However, as this contribution is proportional to the DoS at the energy level under consideration, the contribution vanishes at the nodal ring and the nodal-ring spectrum is not affected. A vanishing DoS implies that screening becomes very weak close to the nodal-ring energy (for a type-I NLSM). The assumption of point-like scatterers should therefore be taken with some reservation, when the disorder originates from charged impurities for example.
		Note that the FB approximation does not properly capture the logarithmic corrections to the self-energy, which can be obtained via the self-consistent Born (SB) approximation \cite{Burkov2011}. This approach yields a finite correction at the nodal-ring energy, but it scales like $\exp[- 2 \pi v^2 / (\gamma Q)]$. A much larger correction (linear in $\gamma$) is obtained in the case of strong tilt, as we will see below. 
		
		In the limit of strong axial tilt $|w_z| \gg |v|$, we can proceed analytically, as in the case of weak tilt, when keeping the leading-order contributions as a function of $v/w_z$ instead of $w_z/v$. The self-energy close to the nodal-ring energy becomes:
		\begin{align} \label{eq:SE_FOB_strong_tilt}
		\begin{split}
		&\Sigma^\FOB(\omega) \approx \frac{\gamma Q}{2 \pi w_z^2} \lef \omega - \imu \, |w_z| k_\cutoff \rig \lef \sigma_0 - \frac{v}{w_z} \, \sigma_3 \rig \\
		&\quad - \frac{\gamma v k_\cutoff \omega}{2 \pi^2 w_z^2} [2 \, \mathrm{arccosh}(|w_z/v|)/w_z + \imu \, \pi/|w_z| ]  \, \sigma_1.
		\end{split}
		\end{align}
		In this case, the imaginary part is cutoff dependent and does not vanish at the nodal-ring energy, unlike the real part. Both the real and the imaginary part vanish in the limit of very large tilt.
		Note that the weak and strong tilt regimes are separated by a Lifshitz transition at $v = w_z$ and that Eqs.~\eqref{eq:SE_FOB_weak_tilt} and \eqref{eq:SE_FOB_strong_tilt} are only valid away from this transition.
		Further note that the FB approximation is only valid when crossing diagrams can safely be neglected (i.e., when $|\gamma k_\cutoff / (2 \pi v w_z)| \ll 1$ in the case of strong tilt) and that it is in agreement with the SB approximation (see Sec.~\ref{sec:SCBA} in Supplemental Material for more details).
		
\begin{figure}[tb]
			\centering
			\includegraphics[width=0.75\linewidth]{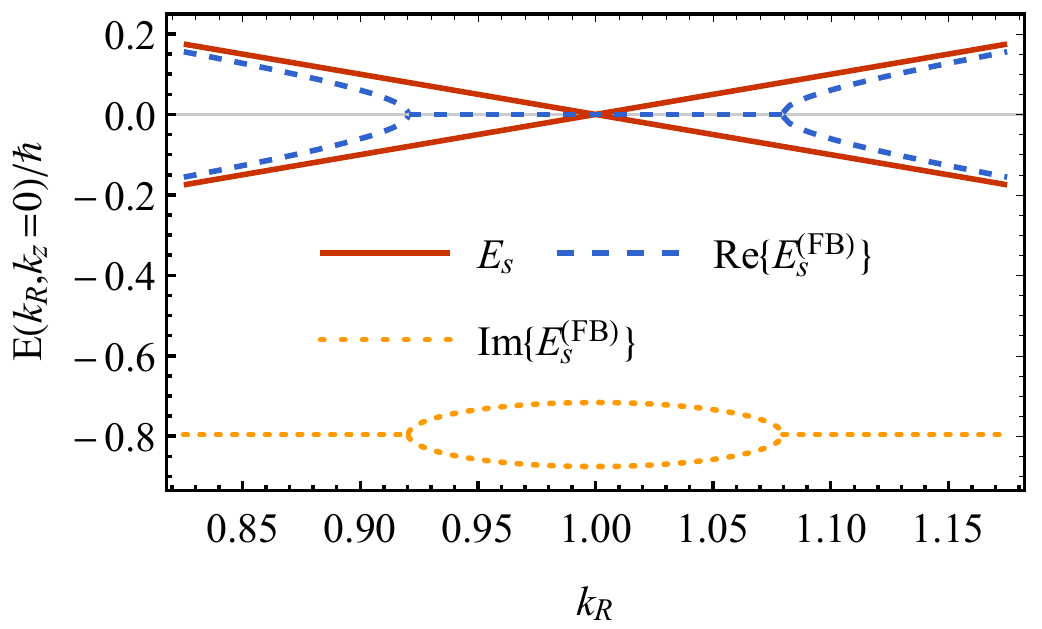}
			\caption{
				The spectrum of a disordered type-II nodal-line semimetal with strong axial tilt (along $k_z$), according to the self-energy correction of Eq.~\eqref{eq:SE_FOB_strong_tilt}, evaluated close to the nodal ring with $k_z = 0$. We have considered the following toy model parameters: $w_z/v = 10$, $\gamma = 100$, $v = 1$, $Q = 1$, $k_\cutoff = 0.5$.
			}
			\label{fig:flat_band}
		\end{figure}

		\textit{Exceptional lines and bulk Fermi ribbons}.---Close to the nodal-ring energy, we can approximate the spectrum by:
			\begin{align} \label{eq:spectrum_FOB_approx}
			\begin{split}
			&E_s^\FOB(\veck)/\hbar \approx \vecw \cdot (\veck - Q \, \vece_{k_R}) + \Sigma_0^\FOB(\omega = 0) \\
			&\quad + s \sqrt{\sum\nolimits_\alpha [\vecv_\alpha \cdot (\veck - Q \, \vece_{k_R}) + \Sigma_\alpha^\FOB(\omega = 0)]^2} \, ,
			\end{split}
			\end{align}
			with $\Sigma^\FOB \equiv \Sigma_\nu^\FOB \, \sigma_\nu$ and summation over $\nu = 0, 1, 2, 3$ according to the Einstein summation convention.
			The nonvanishing imaginary term $\Sigma_3^\FOB(\omega = 0)$ induces a purely imaginary gap in the quasiparticle spectrum for $k_z = 0$ and $k_R$ close to the nodal ring. For the real part of the spectrum, a halo-shaped band touching region develops [see Figs.~\subref*{fig:band_structure_d}, \ref{fig:flat_band}, and \subref*{fig:Fermi_ribbons} (left)], forming a so-called bulk Fermi ribbon. The ribbon width in reciprocal space is approximately equal to $2 / |w_z \tau| \approx 2 / l_R$, with $\tau \approx 2 \pi |w_z| / (\gamma Q k_\cutoff)$ the quasiparticle lifetime and $l_R \approx |w_z| \tau$ the scattering length along the radial direction.
		While electron-hole symmetry persists for the real part of the energy spectrum, it does not for the imaginary part. At the center of the bulk Fermi ribbon, the states have a lifetime that splits equally into $\tau /(1 \pm |v/w_z|)$.
		At the edges of the ribbon, exceptional lines appear for which both the real and the imaginary part of the spectral gap vanish. In the limit of no disorder, these exceptional lines trace back to the original nodal ring.
		At the exceptional lines, there is a square-root singularity for both real and imaginary parts of the spectrum. The Hamiltonian becomes defective and the quasiparticle group velocity diverges. An important remark here is that these singularities are lifted by the self-energy terms $\propto \omega$ in Eq.~\eqref{eq:SE_FOB_strong_tilt}, while having been neglected in Eq.~\eqref{eq:spectrum_FOB_approx} and Fig.~\ref{fig:flat_band}. A perfectly flat bulk Fermi ribbon with exceptional lines is only approximately realized in a physical NLSM system with tilt and disorder, and it will be washed out as the disorder strength, ribbon width, and higher-order corrections grow (see Sec.~\ref{sec:higher_order_corrections} in Supplemental Material). Note that the ribbon width increases when approaching the Lifshitz transition at $v = w_z$ and decreases away from it. Further note that a bulk Fermi ribbon with much smaller but finite width is also expected in the case of weak tilt, induced by the logarithmic self-energy corrections from the SB approximation.
		
		In case of strong radial tilt $w_R$, $\Sigma_1^\FOB(\omega = 0)$ is finite and imaginary [approximately equal to $\gamma v Q k_\cutoff / (\pi w_R |w_R|)$], inducing a purely imaginary gap for $k_R = Q$ close to $k_z = 0$. This leads to the formation of a circular bulk Fermi ribbon that stretches along the axial direction [see Fig.~\subref*{fig:Fermi_ribbons} (top)].
		Analogously, a combination of strong radial and axial tilt will lead to a separation of the nodal ring into two exceptional lines and a bulk Fermi ribbon whose surface lies perpendicular to the tilt direction [see Fig.~\subref*{fig:Fermi_ribbons} (bottom)].
		
		One can also imagine a scenario that leads to a bulk Fermi ribbon with nontrivial topology, with two exceptional loops forming a Hopf link for example (linked together exactly once). This requires the rotational symmetry with respect to the nodal ring to be broken without gapping it out, something for which different scenarios can be envisioned in more material-specific NLSM models and self-energy calculations. A simple demonstration that is compatible with our analytical approach is realized by a twisting tilt vector $\vecw = (w_R \cos^2 k_\phi, w_R \sin k_\phi \cos k_\phi, w_z \sin k_\phi)$ [see Fig.~\ref{fig:Fermi_ribbons} (right)]. The single valuedness of the tilt vector around the nodal ring ensures that the bulk Fermi ribbon can only be an orientable (Seifert) surface with an integer winding number, as is required by general considerations of the zero-energy-gap surface \cite{Carlstrom2018A}.
		
		\textit{Disorder with orbital dependence}.---The disorder potential in Eq.~\eqref{eq:disorder} is diagonal in the orbital space of the two-band Hamiltonian of Eq.~\eqref{eq:nodal_loop_Ham}. In general, a disorder potential can also have a matrix structure to represent some form of orbital dependence, which can be expanded over Pauli matrices: $V(\vecr) = V_\nu(\vecr) \, \sigma_\nu$.
			Until now, we have considered a scalar disorder potential ($S_{1,2,3} = 0$), but sometimes it is important to include an orbital dependence. For example, if the two orbitals can be associated to different atom species in a certain compound, they might be affected differently by specific impurities and their typical positioning. The extreme case in which only one orbital and corresponding band is affected, can then be represented by considering $S_0 = \pm S_3$ ($S_{1,2} = 0$) \cite{Papaj2018}.
		
		Within the lowest-order (LO) approximation, the self-energy is given by $n S_\nu \, \sigma_\nu / \hbar$, leading to a shift of the nodal ring in energy ($S_0 \neq 0$) and reciprocal space ($S_{1,3} \neq 0$) or even a gapped spectrum ($S_2 \neq 0$).
			Orbital-dependent disorder also has an impact on the matrix structure of the self-energy within the FB approximation [given by $\sum_\veck (S_\nu \, \sigma_\nu) \Green^{(0)}(\veck, \omega) (S_{\nu'} \sigma_{\nu'})$], affecting the formation of exceptional lines and bulk Fermi ribbons. For example, in the limit of strong tilt, a bulk Fermi ribbon can be formed due to the following imaginary term, $\Sigma_3^\FOB \propto 2 S_0 S_3 / |w_z|$, rather than the term presented in Eq.~\eqref{eq:SE_FOB_strong_tilt}.
		
		In conclusion, we have studied the impact of disorder on the spectrum of 3D nodal-line semimetals with tilt, considering the lowest-order, first-order Born, and self-consistent Born approximations of the self-energy. Our results show that the nodal ring can shift and deform in energy and momentum space.
		Furthermore, the self-energy in general acquires an imaginary off-diagonal term and renders the matrix structure of the Green function non-Hermitian. When such a term persists at the nodal-ring energy, the nodal ring can split into two exceptional lines, connected by a bulk Fermi ribbon on which the real part of the spectral gap vanishes. This scenario markedly develops in case of strong tilt of type-II, with the width of the bulk Fermi ribbon being proportional to the disorder strength and its surface lying perpendicular to the local tilt orientation. Such a bulk Fermi ribbon is in general a closed and orientable surface, with a possibly nontrivial winding number, which can be realized with a tilt vector that features a nontrivial winding number when traced along the nodal ring. The (non-Hermitian) matrix structure of the Green function can also be manipulated by disorder with an orbital dependence.
		
		These results demonstrate that disordered type-II nodal-line semimetals are very promising for the exploration and verification of non-Hermitian extensions of topological phases of matter. Several observable effects can be expected for these exceptional line and bulk Fermi ribbon states. An important feature in this regard is the change in the quasiparticle spectrum at the nodal-line energy, with the appearance of nondispersive bulk (Fermi ribbon) states. Another important feature of NLSMs in this context is the appearance of drumhead surface states. A first study on the impact of non-Hermiticity on drumhead surface states appeared very recently in Ref.~\cite{Lee2018}, showing that there are significant corrections to the drumhead regions. Experimental signatures of exceptional rings, the flat band touching regions and the drumhead surface states have already been measured in non-Hermitian photonics systems \cite{Zhen2015, Cerjan2018B} and we do not see any fundamental obstructions to perform analogous experimental probing in disordered type-II nodal-line semimetals, based on transport measurements or spectroscopy resolving the bulk or surface state spectrum (e.g., angle-resolved photoemission spectroscopy).
		
		\textit{Acknowledgements}.---K.M.\ and T.L.S.\ acknowledge the support by the National Research Fund Luxembourg with ATTRACT Grant No.\ 7556175 and Edvin Idrisov for the fruitful discussions. A.A.Z.\ acknowledges the support by the Academy of Finland.

		\bibliography{2019_Moors_disorder_arxiv_v2}
	
	%%%%%%%%%% Merge with supplemental materials %%%%%%%%%%
	\widetext
	\clearpage
	\begin{center}
		\textbf{\large Supplemental Material}
	\end{center}
	%%%%%%%%%% Merge with supplemental materials %%%%%%%%%%
	%%%%%%%%%% Prefix a "S" to all equations, figures, tables and reset the counter %%%%%%%%%%
	\setcounter{equation}{0}
	\setcounter{figure}{0}
	\setcounter{table}{0}
	\setcounter{page}{1}
	\setcounter{secnumdepth}{3}
	\makeatletter
	\renewcommand{\theequation}{S\arabic{equation}}
	\renewcommand{\thefigure}{S\arabic{figure}}
	\renewcommand{\bibnumfmt}[1]{[S#1]}
	\renewcommand{\citenumfont}[1]{S#1}
	\renewcommand{\thesection}{S\Roman{section}}
	%%%%%%%%%% Prefix a "S" to all equations, figures, tables and reset the counter %%%%%%%%%%
	
	\section{First-order Born approximation}
	\label{sec:FBA}
	In this section, we provide some more details on the calculation of the self-energy with the FB approximation. Essentially, a summation of $\veck$ over the unperturbed retarded Green function [corresponding to the Hamiltonian in Eq.~\eqref{eq:nodal_loop_Ham}] needs to be calculated and plugged into Eq.~\eqref{eq:SE_FOB}. It can be written as follows:
	\begin{equation} \label{eq:Green_analyt_cont}
	\begin{split}
	\Green^{(0)}(\veck, \omega) &= \frac{1}{2} \sum_s \lef \sigma_0 + s \sum_\alpha \sigma_\alpha \vecv_\alpha \cdot (\veck - Q \, \vece_{k_R}) /
	\sqrt{\sum\nolimits_\beta [\vecv_\beta \cdot (\veck - Q \, \vece_{k_R})]^2} \rig \\
	& \qquad \; \; \; \times \lef \frac{\omega - E_s(\veck)/\hbar}{[\omega - E_s(\veck)/\hbar]^2 + \eta^2}
	- \imu \, \pi \, \delta[\omega - E_s(\veck')/\hbar] \rig
	\equiv \sum_\nu \, [\re\{\Green_\nu^{(0)}(\veck, \omega)\} + \imu \, \im\{\Green_\nu^{(0)}(\veck, \omega)\}] \, \sigma_\nu.
	\end{split}
	\end{equation}
	Making use of the notation introduced on the last line of Eq.~\eqref{eq:Green_analyt_cont}, we obtain for the summation over $\im\{\Green_0^{(0)}(\veck, \omega)\}$ (assuming $v_R = v_z = v$ and $w_R = 0$):
	\begin{equation} \label{eq:Green_imag_0}
	\begin{split}
	\sum_\veck \im\{\Green_0^{(0)}(\veck, \omega)\} &= - \frac{1}{2} \sum_s \pi \frac{\Vol}{(2\pi)^2} \int\limits_0^{+\infty} \! \! \deriv k_R \; k_R \int\limits_{-\infty}^{+\infty} \! \! \deriv k_z \;
	\delta(\omega - w_z k_z - s |v| \sqrt{(k_R - Q)^2 + k_z^2}) \\
	&= - \frac{1}{2} \sum_s \pi \frac{\Vol}{(2\pi)^2} Q \int\limits_{-Q}^{+\infty} \deriv q_R
	\int\limits_{-\infty}^{+\infty} \! \! \deriv k_z \;
	\delta(\omega - w_z k_z - s |v| \sqrt{q_R^2 + k_z^2}),
	\end{split}
	\end{equation}
	with $q_R \equiv k_R - Q$.
	
	In the weak tilt limit ($|w_z/v| \ll 1$) close to the nodal-ring energy ($|\omega/v| \ll Q$) and introducing polar coordinates with radius $k = \sqrt{q_R^2 + k_z^2}$ and polar angle $\theta$ ($k_z = k \cos \theta$), the solution $k^\ast$ of the Dirac-delta function is given by:
	\begin{equation}
	k^\ast = \omega / (s|v| + w_z \cos\theta) \approx s \omega / |v| - w_z \omega \cos\theta / v^2.
	\end{equation}
	Plugging in this solution, we obtain:
	\begin{equation}
	\sum_\veck \im\{\Green_0^{(0)}(\veck, \omega)\} = - \pi \frac{\Vol}{(2\pi)^2} \frac{Q}{2} \int\limits_0^{2\pi} \! \deriv \theta \;
	\frac{|\omega|}{(|v| + s w_z \cos\theta)^2}
	\approx - \Vol \frac{Q |\omega|}{4 v^2}.
	\end{equation}
	Similarly, we retrieve for $\im\{\Green_3^{(0)}(\veck, \omega)\}$:
	\begin{equation}
	\begin{split}
	\sum_\veck \im\{\Green_3^{(0)}(\veck, \omega)\} &= - \frac{1}{2} \sum_s s \pi \frac{\Vol}{(2\pi)^2} \int\limits_{-Q}^{+\infty} \! \! \deriv q_R \; (q_R + Q) \int\limits_{-\infty}^{+\infty} \! \! \deriv k_z \;
	\frac{v k_z}{|v| \sqrt{q_R^2 + k_z^2}} \, \delta(\omega - w_z k_z - s |v| \sqrt{q_R^2 + k_z^2}) \\
	&= - \frac{1}{2} \sum_s s \pi \vartheta(s \omega) \frac{\Vol}{(2\pi)^2} \frac{v Q |\omega|}{|v|} \int\limits_0^{2\pi} \deriv \theta \;
	\frac{\cos\theta}{(|v| + s w_z \cos\theta)^2}
	\approx \Vol \frac{Q w_z |\omega|}{4 v^3} \\
	&= - \frac{w_z}{v} \sum_\veck \im\{\Green_0^{(0)}(\veck, \omega)\}.
	\end{split}
	\end{equation}
	Furthermore, we obtain in a similar manner that $\im\{\Green_1^{(0)}(\veck, \omega)\} \propto \omega^2$ is negligible with respect to the other contributions, while $\Green_2^{(0)} = 0$ by definition.
	
	For the real part of the Green function, we get the following contributions:
	\begin{equation} \label{eq:calc_Re_SE_sig0_NLSM_I_axial_tilt}
	\begin{split}
	\sum_\veck \re\{\Green_0^{(0)}(\veck, \omega)\} &= \frac{1}{2} \sum_s \frac{\Vol}{(2\pi)^2} \mathcal{P} \! \int\limits_{-q_R}^{+\infty} \! \! \deriv q_R \; (q_R + Q) \int\limits_{-\infty}^{+\infty} \! \! \deriv k_z \;
	\frac{1}{\omega - w_z k_z - s |v| \sqrt{q_R^2 + k_z^2}} \\
	&\rightarrow \frac{1}{2} \sum_s \frac{\Vol}{(2\pi)^2} \mathcal{P} \, Q \int\limits_0^{2\pi} \deriv \theta \int\limits_0^{k_\cutoff^R} \deriv k \;
	\frac{k}{\omega - (s|v| + w_z \cos\theta) k}
	\approx -\Vol \frac{Q \omega}{2 \pi v^2} \ln\left|\frac{v k_\cutoff}{\omega}\right|,
	\end{split}
	\end{equation}
	where the arrow indicates the introduction of a cutoff wavevector $k_\cutoff$ and we only keep the terms up to leading order in $k_\cutoff$. The cutoff wave vector represents the maximal distance to the nodal ring for the integration over $\veck$ in both the radial and the axial directions. The energy scale of $\omega$ and the cutoff wave vector are assumed to obey the following constraints: $|\omega/v| \ll k_\cutoff < Q$.
	Similarly, we obtain:
	\begin{equation} \label{eq:calc_Re_SE_sig1_NLSM_I_axial_tilt}
	\begin{split}
	\sum_\veck \re\{\Green_1^{(0)}(\veck, \omega)\} &= \frac{1}{2} \sum_s s \frac{\Vol}{(2\pi)^2} \mathcal{P} \,
	\int\limits_{-Q}^{+\infty} \! \! \deriv q_R \; (q_R + Q) \int\limits_{-\infty}^{+\infty} \! \! \deriv k_z \;
	\frac{v q_R}{|v| \sqrt{q_R^2 + k_z^2}} \, \frac{1}{\omega - w_z k_z - s |v| \sqrt{q_R^2 + k_z^2}} \\
	&\rightarrow \frac{1}{2} \sum_s s \frac{\Vol}{(2\pi)^2} \frac{2 v}{|v|} \mathcal{P} \,
	\int\limits_0^\pi \deriv \theta \, \sin^2\!\theta \int\limits_0^{k_\cutoff} \deriv k \;
	\frac{k^2}{\omega - (s|v| + w_z \cos\theta) k} \approx - \Vol \frac{k_\cutoff^2}{8 \pi v},
	\end{split}
	\end{equation}
	having considered 4D spherical coordinates on the last line, and
	\begin{equation} \label{eq:calc_Re_SE_sig3_NLSM_I_axial_tilt}
	\begin{split}
	\sum_\veck \re\{\Green_3^{(0)}(\veck, \omega)\} &\rightarrow \frac{1}{2} \sum_s s \frac{\Vol}{(2\pi)^2} \mathcal{P} \, \frac{v Q}{|v|} \int\limits_0^{2 \pi} \deriv \theta \; \cos\theta \int\limits_0^{k_\cutoff} \deriv k \;
	\frac{k}{\omega - (s|v| + w_z \cos\theta) k} \\
	&\approx \Vol \frac{w_z Q \omega}{2 \pi v^3} \ln\left| \frac{v k_\cutoff}{\omega} \right|
	= - \frac{w_z}{v} \sum_\veck \re\{\Sigma_0^\FOB(\omega)\}.
	\end{split}
	\end{equation}
	Combining all these results yields Eq.~\eqref{eq:SE_FOB_weak_tilt}. Note that the angular dependence of the integrands in Eqs.~\eqref{eq:calc_Re_SE_sig0_NLSM_I_axial_tilt}-\eqref{eq:calc_Re_SE_sig3_NLSM_I_axial_tilt} is different. The $\sigma_0$-term diverges in all directions, whereas the $\sigma_{3(1)}$ term only diverges away from the radial plane (axial direction). If an anisotropic cutoff is considered, the connection between Eq.~\eqref{eq:calc_Re_SE_sig0_NLSM_I_axial_tilt} and Eq.~\eqref{eq:calc_Re_SE_sig3_NLSM_I_axial_tilt} (last equality) would not hold for example. The crucial aspects for the phenomenology of exceptional lines and bulk Fermi ribbons, however, is the proportionality of self-energy terms to the Pauli matrices and their dependency on $\omega$ and they do not depend on the details of the cutoff implementation. Similar remarks can be made for the self-energy terms in the case of strong tilt.
	
	In case of very strong tilt $|w_z/v| \gg 1$, the Dirac-delta function in Eq.~\eqref{eq:Green_imag_0} has solutions within the integration boundaries when
	\begin{equation} \label{eq:k_min}
	-1 \leq \frac{\omega - s |v| k}{w_z k} \leq 1 \quad \Rightarrow \quad k \geq \frac{- s |v| \omega + |w_z \omega|}{w_z^2 - v^2} \equiv k_{\min}.
	\end{equation}
	This implies that we have to introduce a cutoff wavevector for the imaginary part as well, unlike in the case of weak tilt. Integrating over $\theta$ first and assuming an isotropic cutoff, we obtain:
	\begin{equation}
	\begin{split}
	\sum_\veck \im\{\Green_0^{(0)}(\veck, \omega)\} &\rightarrow - \frac{1}{2} \sum_s \pi \frac{\Vol}{(2\pi)^2} \, 2 Q \! \int\limits_{k_{\min}}^{k_\cutoff} \! \! \deriv k \;
	\frac{k}{\sqrt{(w_z k)^2 - (\omega - s |v| k)^2}}
	\approx - \Vol \frac{Q k_\cutoff}{2 \pi |w_z|}, \\
	\sum_\veck \im\{\Green_1^{(0)}(\veck, \omega)\} &\rightarrow - \frac{1}{2} \sum_s s \pi \frac{\Vol}{(2\pi)^2} \frac{2 v}{|v|} \int\limits_0^\pi \deriv \theta \; \sin^2\!\theta \int\limits_0^{k_\cutoff} \deriv k \;
	k^2 \, \delta(\omega - w_z k \cos\theta - s |v| k) \\
	&\approx - \frac{1}{2} \sum_s \pi \frac{\Vol}{(2\pi)^2} \frac{2 v \omega}{w_z^2} \int\limits_{k_{\min}}^{k_\cutoff} \! \! \deriv k \;
	\frac{k}{\sqrt{(w_z k)^2 - \omega^2}}
	\approx - \Vol \frac{v k_\cutoff \omega}{2 \pi |w_z|^3}, \\
	\sum_\veck \im\{\Green_3^{(0)}(\veck, \omega)\} &\rightarrow - \frac{1}{2} \sum_s s \pi \frac{\Vol}{(2\pi)^2} \frac{v Q}{|v|} \int\limits_0^{2\pi} \deriv \theta \; \cos\theta \int\limits_0^{k_\cutoff} \deriv k \;
	k \, \delta(\omega - w_z k \cos\theta - s |v| k) \\
	&= - \frac{1}{2} \sum_s s \pi \frac{\Vol}{(2\pi)^2} \frac{2 v Q}{|v| w_z} \int\limits_{k_{\min}}^{k_\cutoff} \! \! \deriv k \;
	\frac{\omega - s|v| k}{\sqrt{(w_z k)^2 - (\omega - s |v| k)^2}}
	\approx \Vol \frac{v Q k_\cutoff}{2 \pi w_z |w_z|} \\
	&= - \frac{v}{w_z} \sum_\veck \im\{\Green_0^{(0)}(\veck, \omega)\}.
	\end{split}
	\end{equation}

	The summation over the real part of the Green function yields the following components:
	\begin{equation}
	\begin{split}
	\sum_\veck \re\{\Green_0^{(0)}(\veck, \omega)\} &\rightarrow \frac{1}{2} \sum_s \frac{\Vol}{(2 \pi)^2} \, 4 Q \, \mathcal{P} \int\limits_0^{k_\cutoff} \deriv k \; \int\limits_0^1 \deriv x \;
	\frac{1}{\sqrt{1 - x^2}} \frac{k (\omega - s|v| k)}{(\omega - s |v| k)^2 - (w_z k)^2 x^2} \\
	&= \frac{1}{2} \sum_s \frac{\Vol}{(2 \pi)^2} \, 2 \pi Q \! \int\limits_0^{k_{\max}} \! \! \deriv k \;
	\frac{k (\omega - s |v| k)}{|\omega - s |v| k| \sqrt{(\omega - s |v| k)^2 - (w_z k)^2}} \\
	&\approx \frac{\Vol}{(2 \pi)^2} \frac{2 \pi Q \omega}{|\omega|} \! \! \int\limits_0^{|\omega/w_z|} \! \! \! \! \deriv k \; 
	\frac{k}{\sqrt{\omega^2 - w_z^2 k^2}}
	= \Vol \frac{Q \omega}{2 \pi w_z^2},
	\end{split}
	\end{equation}
	with $x \equiv \cos\theta$, $k_{\max} \equiv k_{\min}$ from Eq.~\eqref{eq:k_min}, and where we have made use of the following identity:
	\begin{equation}
	\int\limits_0^1 \deriv x \; \frac{1}{\sqrt{1 - x^2}} \frac{1}{a^2 - b^2 x^2} = \left\{ \begin{matrix*}[l]
	-\imu \pi / (2 |a| \sqrt{b^2 - a^2}) & \quad (|b| > |a| > 0) \\
	\pi / (2 |a| \sqrt{a^2 - b^2}) & \quad (|a| > |b| > 0)
	\end{matrix*} \right.,
	\end{equation}
	for $a, b \in \mathbb{R}$, and
	\begin{equation}
	\begin{split}
	\sum_\veck \re\{\Green_1^{(0)}(\veck, \omega)\} &\rightarrow \frac{1}{2} \sum_s s \frac{\Vol}{(2\pi)^2} \frac{4 v w_z}{|v|} \mathcal{P} \int\limits_0^{k_\cutoff} \deriv k \int\limits_0^1 \deriv x \;
	\frac{k^3 \sqrt{1 - x^2} x}{(\omega - s |v| k)^2 - (w_z k)^2 x^2} \\
	&\approx - \frac{1}{2} \sum_s s \frac{\Vol}{(2\pi)^2} \frac{4 v w_z}{|v w_z^3|} \int\limits_{k_{\min}}^{k_\cutoff} \! \! \deriv k \;
	\sqrt{(w_z k)^2 - (\omega - s |v| k)^2} \, \mathrm{arccosh}(|w_z k|/|\omega - s |v| k|) \\
	&\approx - \Vol \frac{v k_\cutoff \omega}{\pi^2 w_z^3} \, \mathrm{arccosh}(|w_z/v|),
	\end{split}
	\end{equation}
	where we have made use of the following identity:
	\begin{equation}
	\int\limits_0^1 \deriv x \; \frac{\sqrt{1 - x^2} x}{a^2 - b^2 x^2} = \left\{ \begin{matrix*}[l]
	- \imu \pi \sqrt{b^2 - a^2} / (2 |b|^3) + 1/b^2 - \sqrt{b^2 - a^2} \, \mathrm{arccosh}(|b/a|) / |b|^3  & \quad (|b| > |a| > 0) \\
	1/b^2 - \sqrt{a^2 - b^2} \arcsin(|b/a|) / |b|^3 & \quad (|a| > |b| > 0)
	\end{matrix*} \right.,
	\end{equation}
	and
	\begin{equation}
	\begin{split}
	\sum_\veck \re\{\Green_3^{(0)}(\veck, \omega)\} &\rightarrow \frac{1}{2} \sum_s s \frac{\Vol}{(2 \pi)^2} \frac{4 v w_z Q}{|v|} \mathcal{P} \int\limits_0^{k_\cutoff} \deriv k \int\limits_0^1 \deriv x \;
	\frac{1}{\sqrt{1 - x^2}} \frac{k^2 x^2}{(\omega - s |v| k)^2 - (w_z k)^2 x^2} \\
	&= \frac{1}{2} \sum_s s \frac{\Vol}{(2 \pi)^2} \frac{4 \pi v Q}{2 |v| w_z} \mathcal{P} \int\limits_0^{k_{\max}} \! \! \deriv k \;
	\frac{|\omega - s |v| k|}{\sqrt{(\omega - s |v| k)^2 - (w_z k)^2}} \\
	&\approx - \frac{\Vol}{(2 \pi)^2} \frac{2 \pi v Q \omega}{w_z |\omega|} \int\limits_0^{|\omega/w_z|} \! \! \! \! \deriv k \;
	\frac{k}{\sqrt{\omega^2 - w_z^2 k^2}}
	= - \Vol \frac{v Q \omega}{2 \pi w_z^3} = - \frac{v}{w_z} \sum_\veck \re\{\Green_0^{(0)}(\veck, \omega)\},
	\end{split}
	\end{equation}
	where we have made use of the following identity:
	\begin{equation}
	\int\limits_0^1 \deriv x \; \frac{1}{\sqrt{1 - x^2}} \frac{x^2}{a^2 - b^2 x^2} = \left\{ \begin{matrix*}[l]
	- \pi [ 1 + \imu |a| /\sqrt{b^2 - a^2}] / (2 b^2) & \quad (|b| > |a| > 0) \\
	- \pi [ 1 - |a| /\sqrt{a^2 - b^2}] / (2 b^2) & \quad (|a| > |b| > 0)
	\end{matrix*} \right. .
	\end{equation}
	Combining all these results yields Eq.~\eqref{eq:SE_FOB_strong_tilt}.
	
	It is a straightforward calculation to verify that the energy-independent imaginary contribution to the summation over the Green function in case of very strong radial tilt ($|w_R/v| \gg 1$) is proportional to $\sigma_1$ rather than $\sigma_3$ in case of strong axial tilt. Following the same procedure as for axial tilt, we get:
	\begin{equation}
	\begin{split}
	\sum_\veck \im\{\Green_0^{(0)}(\veck, \omega)\} &= \frac{1}{2} \sum_s \pi \frac{\Vol}{(2\pi)^2} \int\limits_{-Q}^{+\infty} \deriv q_R \; (q_R + Q)
	\int\limits_{-\infty}^{+\infty} \! \! \deriv k_z \;
	\delta(\omega - w_R q_R - s |v| \sqrt{q_R^2 + k_z^2}) \\
	&= \underbrace{-\frac{1}{2} \sum_s \pi \frac{\Vol}{(2\pi)^2} \int\limits_0^\pi \deriv \theta \; \sin\theta \int\limits_0^{+\infty} \deriv k \; k^2 \,
		\delta(\omega - w_R k \sin\theta - s |v| k)}_{(\mathrm{A})} - \underbrace{(w_R \leftrightarrow -w_R)}_{(\mathrm{B})} \\
	&\quad + \underbrace{-\frac{1}{2} \sum_s \pi \frac{\Vol}{(2\pi)^2} Q \int\limits_0^{2\pi} \deriv \theta \int\limits_0^{+\infty} \deriv k \; k \,
		\delta(\omega - w_R k \sin\theta - s |v| k)}_{(\mathrm{C})},
	\end{split}
	\end{equation}
	with the terms evaluating as follows:
	\begin{equation}
	\begin{split}
	(\mathrm{A}) &\rightarrow -\frac{1}{2} \sum_s \pi \frac{\Vol}{(2\pi)^2} \frac{2}{w_R} \int\limits_{k_{\min}}^{k_\cutoff} \! \! \deriv k \;
	\frac{k (\omega - s|v| k)}{\sqrt{(w_R k)^2 - (\omega - s |v| k)}}
	\approx -\Vol \frac{k_\cutoff \omega}{2 \pi w_R |w_R|}, \\
	(\mathrm{C}) &\rightarrow -\frac{1}{2} \sum_s \pi \frac{\Vol}{(2\pi)^2} 4 Q \int\limits_{k_{\min}}^{k_\cutoff} \! \! \deriv k \;
	\frac{k}{\sqrt{(w_R k)^2 - (\omega - s |v| k)}}
	\approx -\Vol \frac{Q k_\cutoff}{\pi |w_R|},
	\end{split}
	\end{equation}
	resulting in
	\begin{equation}
	\sum_\veck \im\{\Green_0^{(0)}(\veck, \omega)\} \approx -\Vol \frac{Q k_\cutoff}{\pi |w_R|}.
	\end{equation}
	Similarly, we obtain:
	\begin{equation}
	\begin{split}
	\sum_\veck \im\{\Green_1^{(0)}(\veck, \omega)\} &= -\frac{1}{2} \sum_s s \pi \frac{\Vol}{(2\pi)^2} \frac{v}{|v|} \int\limits_{-Q}^{+\infty} \deriv q_R \; (q_R + Q)
	\int\limits_{-\infty}^{+\infty} \! \! \deriv k_z \;
	\frac{q_R}{\sqrt{q_R^2 + k_z^2}}
	\delta(\omega - w_R q_R - s |v| \sqrt{q_R^2 + k_z^2}) \\
	&= \underbrace{-\frac{1}{2} \sum_s s \pi \frac{\Vol}{(2\pi)^2} \frac{v}{|v|}
		\int\limits_0^\pi \deriv \theta \; \sin^2\theta
		\int\limits_0^{+\infty} \deriv k \; k^2 \,
		\delta\lef \omega - w_R k \sin\theta - s |v| k \rig}_{(\mathrm{D})} + \underbrace{(w_R \leftrightarrow -w_R)}_{(\mathrm{E})} \\
	&\quad + \underbrace{-\frac{1}{2} \sum_s s \pi \frac{\Vol}{(2\pi)^2} \frac{v Q}{|v|} \int\limits_0^\pi \deriv \theta \; \sin\theta \int\limits_0^{+\infty} \deriv k \;
		k \, \delta\lef \omega - w_R k \sin\theta - s |v| k \rig}_{(\mathrm{F})} - \underbrace{(w_R \leftrightarrow -w_R)}_{(\mathrm{G})},
	\end{split}
	\end{equation}
	with
	\begin{equation}
	\begin{split}
	(\mathrm{D}) &\rightarrow -\frac{1}{2} \sum_s s \pi \frac{\Vol}{(2\pi)^2} \frac{2 v}{|v| w_R^2} \int\limits_{k_{\min}}^{k_\cutoff} \! \! \deriv k \;
	\frac{(\omega - s |v| k)^2}{\sqrt{(w_R k)^2 - (\omega - s |v| k)^2}}
	\approx \Vol \frac{v k_\cutoff \omega}{\pi |w_R|^3}, \\
	(\mathrm{F}) &\rightarrow -\frac{1}{2} \sum_s s \pi \frac{\Vol}{(2\pi)^2} \frac{2 v Q}{|v| w_R} \int\limits_{k_{\min}}^{k_\cutoff} \! \! \deriv k \;
	\frac{\omega - s |v| k}{\sqrt{(w_R k)^2 - (\omega - s |v| k)^2}}
	\approx \Vol \frac{v Q k_\cutoff}{2 \pi w_R |w_R|},
	\end{split}
	\end{equation}
	leading to
	\begin{equation}
	\sum_\veck \im\{\Green_1^{(0)}(\veck, \omega)\} \approx \Vol \frac{v Q k_\cutoff}{\pi w_R |w_R|} = - \frac{v}{w_z} \sum_\veck \im\{\Green_0^{(0)}(\veck, \omega)\},
	\end{equation}
	and finally
	\begin{equation}
	\begin{split}
	\sum_\veck \im\{\Green_3^{(0)}(\veck, \omega)\} &= -\frac{1}{2} s \pi \frac{\Vol}{(2\pi)^2} \frac{v}{|v|} \int\limits_{-Q}^{+\infty} \deriv q_R \; (q_R + Q)
	\int\limits_{-\infty}^{+\infty} \! \! \deriv k_z \;
	\frac{k_z}{\sqrt{q_R^2 + k_z^2}}
	\delta(\omega - w_R q_R - s |v| \sqrt{q_R^2 + k_z^2}) = 0.
	\end{split}
	\end{equation}

	\section{Self-consistent Born approximation}
	\label{sec:SCBA}
	The SB approximation, considering a scalar disorder potential ($S_{1, 2, 3} = 0$) and energies close to the nodal loop ($\omega \rightarrow 0$), leads to the following self-consistency equation for the retarded self-energy $\Sigma^\SCB \equiv \lim_{\omega\rightarrow 0} \Sigma^\SCB(\omega)$:
	\begin{align} \label{eq:SCB}
	\Sigma^\SCB = \frac{\gamma}{(2 \pi)^3} \int \deriv^3 k \; \frac{1}{\omega - H(\veck)/\hbar - \Sigma^\SCB}.
	\end{align}
	We consider the following Hamiltonian for a NLSM with nodal ring with radius $Q$ in the $k_z = 0$ plane, positive velocities $v_R = v_z = v > 0$ and positive axial tilt ($w_R = 0$, $w_z > 0$):
	\begin{align}
	H(\veck)/\hbar = w_z k_z \, \sigma_0 + (v Q / 2) [(k_R/Q)^2 - 1] \, \sigma_1 + v k_z \, \sigma_3.
	\end{align}
	Note that this Hamiltonian has a quadratic term, which vanishes close to the nodal loop and regularizes the nonphysical cone tops of the linearized model, present in Eq.~\eqref{eq:nodal_loop_Ham}.
	Solving for $\Sigma^\SCB$ in Eq.~\eqref{eq:SCB}, we get:
	\begin{equation}
	\begin{split}
	\Sigma_0^\SCB &= \frac{\gamma}{2 \pi} \left[ \lef -\frac{\omega - \re\{\Sigma_0^\SCB\} + c_z \re\{\Sigma_3^\SCB\}}{c_z^2 - 1} - \imu \, \im\{\Sigma_0^\SCB\} \rig \int \! \deriv k \; J - c_z \int \! \deriv k \; v k J \right], \\
	\Sigma_3^\SCB &= \frac{\gamma}{2 \pi} \left[ \lef c_z \frac{\omega - \re\{\Sigma_0^\SCB\} + c_z \re\{\Sigma_3^\SCB\}}{c_z^2 - 1} + \imu \, \im\{\Sigma_3^\SCB\} \rig \int \! \deriv k \; J + \int \! \deriv k \; v k J \right],
	\end{split}
	\end{equation}
	with $c_z \equiv w_z/v$,
	\begin{equation}
	\begin{split}
	J &\equiv \frac{1}{2 \pi} \int \! \deriv k_R \; k_R \det (k_R^2 / Q^2)^{-1}, \\
	\det(x) &\equiv (\omega - w_z k_z - \Sigma_0^\SCB)^2 - (v Q/2)^2 (x - 1)^2 - (v k_z + \Sigma_3^\SCB)^2,
	\end{split}
	\end{equation}
	and $k \equiv k_z + q$, with:
	\begin{equation}
	q \equiv \frac{1}{v} \frac{c_z (\omega - \re\{\Sigma_0^\SCB\}) + \re\{\Sigma_3^\SCB\}}{c_z^2 - 1}.
	\end{equation}

	For $c_z > 1$, we get the following solutions for the integration over $J$ and $v k J$, respectively:
	\begin{equation}
	\begin{split}
	-I_1 &\equiv \frac{\gamma}{2\pi} \int \! \deriv k \; J \approx -\frac{\gamma Q}{4 v^2 \sqrt{c_z^2 - 1}}, \\
	\imu I_2 &\equiv \frac{\gamma}{2 \pi} \int \! \deriv k \; v k J \approx \imu \frac{\gamma Q k_\cutoff}{2 \pi v \sqrt{c_z^2 - 1}} \sign(- c_z \im\{\Sigma_0^\SCB\} + \im\{\Sigma_3^\SCB\}),
	\end{split}
	\end{equation}
	with cutoff wave vector $k_\cutoff$ for the integration over $k$ along the axial direction (note that this cutoff implementation differs from that of the FB approximation, but that it does not affect the phenomenology of the quasiparticle spectrum).
	The newly introduced integration constants, $I_1$ and $I_2$, are approximately real, such that the self-energy becomes:
	\begin{equation}
	\begin{split}
	\Sigma_0^\SCB &= \frac{I_1}{c_z^2 (1 + I_1) + I_1 - 1} \omega - \imu \frac{c_z I_2}{1 - I_1}, \\
	\Sigma_3^\SCB &= - \frac{c_z I_1}{c_z^2 (1 + I_1) + I_1 - 1} \omega + \imu \frac{I_2}{1 + I_1} = - c_z \re\{\Sigma_0^\SCB\} - \imu \, \im\{\Sigma_0^\SCB\} / c_z.
	\end{split}
	\end{equation}
	In the limit $c_z \gg 1$, $I_{1, 2} \ll 1$ we obtain:
	\begin{equation}
	\begin{split}
	\Sigma_0^\SCB &\approx \imu \, \gamma \frac{Q k_\cutoff}{2 \pi w_z}, \\
	\Sigma_3^\SCB &\approx - \imu \, \gamma \frac{v Q k_\cutoff}{2 \pi w_z^2},
	\end{split}
	\end{equation}
	recovering the result of Eq.~\eqref{eq:SE_FOB_strong_tilt} that is responsible for the appearance of a bulk Fermi ribbon.
	
	For $c_z < 1$, the integrations over $J$ and $v k J$ yield:
	\begin{equation}
	\begin{split}
	\frac{\gamma}{2\pi} \int \! \deriv k \; J &\approx -\frac{\gamma Q}{2 \pi v^2 \sqrt{1 - c_z^2}} \ln \left| \frac{k_\cutoff}{q} \right| - \imu \frac{\gamma Q \, \sign(\Gamma)}{4 v^2 \sqrt{1 - c_z^2}} \equiv - H_1 - \imu H_2, \\
	\frac{\gamma}{2 \pi} \int \! \deriv k \; v k J &\approx - \frac{\gamma Q [c_z (\omega - \re\{\Sigma_0^\SCB\}) + \re\{\Sigma_3^\SCB\}]}{2 \pi v^2 (1 - c_z^2)^{3/2}} \equiv -H_3 [c_z (\omega - \re\{\Sigma_0^\SCB\}) + \re\{\Sigma_3^\SCB\}],
	\end{split}
	\end{equation}
	with:
	\begin{equation}
	\Gamma \equiv (\omega - \re\{\Sigma_0^\SCB\} + c_z \re\{\Sigma_3^\SCB\}) (-\im\{\Sigma_0^\SCB\} + c_z \im\{\Sigma_3^\SCB\}).
	\end{equation}
	In the limit $c_z \ll 1$, $H_{1, 2, 3} \ll 1$, we obtain:
	\begin{equation}
	\begin{split}
	\Sigma_0^\SCB &\approx - H_1 \omega - \imu \, H_2 \omega \approx -\frac{\gamma Q \omega}{2 \pi v^2} \ln \left| \frac{k_\cutoff}{q} \right| - \imu \, \frac{\gamma Q |\omega|}{4 v^2}, \\
	\Sigma_3^\SCB &\approx c_z H_1 \omega + \imu \, c_z H_2 \omega = - c_z \Sigma_0^\SCB.
	\end{split}
	\end{equation}
	Up to a cutoff-independent constant in the real part, this is in exact agreement with the result based on the FB approximation in Eq.~\eqref{eq:SE_FOB_weak_tilt}. Note that we have neglected the $\propto \sigma_1$ self-energy correction that leads to a renormalization of the nodal-ring radius in this section, as it is irrelevant for the appearance of a bulk Fermi ribbon.
	
	\section{Higher-order corrections}
	\label{sec:higher_order_corrections}
\begin{figure}[tb]
		\centering
		\includegraphics[width=0.35\linewidth]{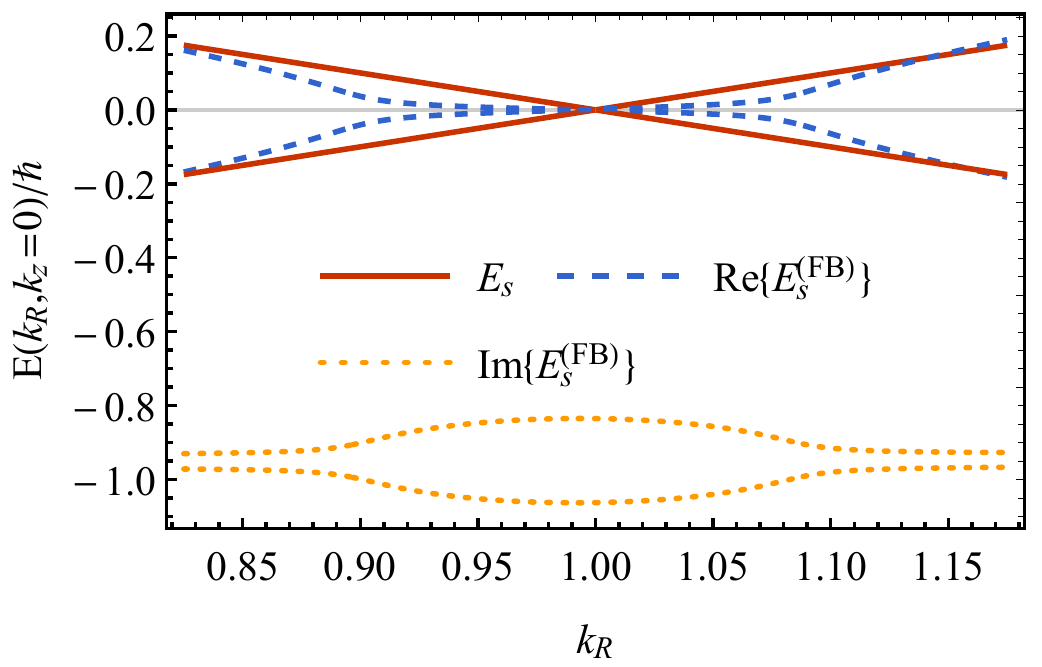}
		\caption{
			The spectrum of a type-II nodal-line semimetal with strong axial tilt (along $k_z$) and disorder, according to the solution of Eq.~\eqref{eq:spectrum_FOB_strong_tilt}, evaluated close to the nodal ring with $k_z = 0$. The same parameter set as in Fig.~\ref{fig:flat_band} has been considered.
		}
		\label{fig:flat_band_corrections}
	\end{figure}

	In Fig.~\ref{fig:flat_band}, we have plotted the complex quasiparticle spectrum for strong axial tilt while neglecting the $\propto \omega$ terms in the self-energy, resulting in a flat bulk Fermi ribbon with a sharp square-root singularity at its edges.
	We can also solve for the spectrum while taking into account the higher-order corrections.
	From Eq.~\eqref{eq:SE_FOB_strong_tilt}, we can write the self-energy as follows:
	\begin{equation}
	\Sigma(\omega) = (\Theta \omega - \imu / \tau) (\sigma_0 - \sigma_3/c_z) + (\Xi \omega + \imu \Phi \omega) \, \sigma_1,
	\end{equation}
	with real parameters $\Theta, 1/\tau, \Xi$ and $\Phi$, all being proportional to $\gamma$.
	Plugging this into Eq.~\eqref{eq:SE_FOB} and solving for $E_s(\veck)$ with Eq.~\eqref{eq:SE_FOB_eq}, yields:
	\begin{equation} \label{eq:spectrum_FOB_strong_tilt}
	\begin{split}
	E_s(\veck)/\hbar &= \frac{g(k_R, k_z) + s \sqrt{g(k_R, k_z)^2 + f h(k_R, k_z)}}{f}, \qquad f \equiv (1 - \Theta)^2 - \Theta^2/c_z^2 - (\Xi + \imu \, \Phi)^2, \\
	g(k_R, k_z) &\equiv [1 - (1-1/c_z^2)\Theta](w_z k_z - \imu/\tau) + v(k_R - Q)(\Xi + \imu \Phi), \\
	h(k_R, k_z) &\equiv v^2 (k_R - Q)^2 - (1 - 1/c_z^2) (w_z k_z - \imu / \tau)^2.
	\end{split}
	\end{equation}
	A flat bulk Fermi ribbon is realized for values of $k_R$ and $k_z$ that satisfy:
	\begin{equation}
	g(k_R, k_z) = C, \qquad
	0 > g(k_R, k_z)^2 + f h(k_R, k_z) \in \mathbb{R},
	\end{equation}
	with constant $C$. Only when neglecting $\Theta, \Xi$ and $\Phi$ is this realized by $k_z = 0$ and $|k_R - Q| < 1/|w_z \tau|$. In general, these conditions cannot be met. In Fig.~\ref{fig:flat_band_corrections}, the exact spectrum, according to Eq.~\eqref{eq:spectrum_FOB_strong_tilt} is presented for the same parameters as considered in Fig.~\ref{fig:flat_band_corrections}, with the flat bulk Fermi ribbon and square-root singularities only approximately realized.
	
\end{document}